\documentstyle[12pt]{article}
\def\dis{\displaystyle}
\newcommand{\ba}{\mbox{\bf {$\alpha$~}}}
\newcommand{\bb}{\mbox{\bf {$\beta$~}}}
\def\l{\left}
\def\r{\right}
\def\be{\begin{equation}}
\def\ee{\end{equation}}
\def\barr{\begin{eqnarray}}
\def\earr{\end{eqnarray}}
\def\lan{\langle}
\def\ran{\rangle}
\begin{document}    
\begin{flushright}
\end{flushright}
\vskip 5pt
\begin{center}
{\Large {\bf Half lives and Rates for Electron Capture on Drip Line
Nuclei and for Beta Decay with $65 < A <100$
}}
\vskip 20pt
\renewcommand{\thefootnote}{\fnsymbol{footnote}}
 
Debasish Majumdar 
\footnote{E-mail address: debasish@theory.saha.ernet.in}    
and Kamales Kar
\footnote{E-mail address: kamales@theory.saha.ernet.in}
\vskip 10pt
{\it Theory Division, Saha Institute of Nuclear Physics,\\
1/AF Bidhannagar, Kolkata 700064, India. }\\  
\end{center}
\begin{center}
{\bf Abstract}
\end{center}

{\small
\noindent The half lives are calculated for the process of $\beta^\pm$ decay 
and electron capture for nuclei in  mass range $\sim$ 65 - 100 
relevant for the core of a massive star  at late burning stage of steller 
evolution that 
leads to supernova explosion. The rates of electron capture process    
are also computed for nuclei relevant during presupernova stage
of massive stars. These half lives and rates are calculated   
by expressing $\beta^\pm$ Gamow-Teller decay strengths in terms of smoothed 
bivariate strength densities.  These strength
densities are constructed in the framework of spectral averaging theory for 
two body nuclear Hamiltonian in a large nuclear shell model space. 
The method has a natural extension to weak interaction rates for r 
and rp-processes.
}

\section{Introduction}

The late evolution stages of massive stars ($>8M_\odot$) is strongly 
influenced by the weak interaction processes.  
In the pre-collapse stage of a star  the important 
weak processes are electron captures and beta decays. These weak processes 
influence the value of $Y_e$, the electron fraction. As Chandrasekhar 
mass is proportional to $Y_e^2$, the value of $Y_e$ is crucial for the 
collapse of a supernova star and the subsequent evolution. 
The stability (or collapse) of the star
at late stage is determined by among other things, the balancing of the 
gravitational collapse by electron degenerate pressure. While electron 
capture reduces the number of electrons for pressure support, $\beta^-$ 
decays increases the same. Therefore these rates are important ingredients
for building a model for the core collapse supernova. 
With this in view, in this work 
we have investigated the rates and half lives for weak interaction processes
for some of the nuclei relevant for presupernova and supernova environment
with $65<A<100$. 
We first calculate the weak interaction strength densities using a theory 
based on principles of statistical nuclear spectroscopy -- spectral averaging 
theory. 

Among the earlier calculations for weak interaction rates ($\beta^\pm$, 
EC etc.), Fuller, Fowler, Newman \cite{ffn} for the first time dealt with 
these rates at finite temperature in the mass range $21 \le A \le 60$. 
Aufderhide {\em et al} \cite{au,au2} emphasised the need for calculating
weak interaction rates for $A >60$ nuclei as presupernova stars generate
neutron rich cores with mass $A > 60$. Aufderhide et al and later 
Kar et al \cite{kar0} have made rate calculations in the mass 
$A > 60$. In this work, we have explicitly constructed the bivariate
weak interaction strength densities for calculation of weak interaction 
half lives and rates. As in an earlier work, using these methods, $\beta^-$ 
decay rates have been calculated \cite {majumdar4} for nuclei in the regime 
$55 < A < 65$ and extended to EC for a few fp-shell nuclei, 
in this work we have computed  EC half lives for the nuclear mass region
65 - 81 that are still relevant for supernova stars using smoothed
form for EC strength densities in the framework of   
spectral averaging theory. We also calculate rates for some sample 
nuclei in this region and this would be extended to neutron rich nuclei
later. 
 
After the spontaneous nucleosynthesis ceases in late stage of steller 
evolution, heavier elements in the star
are synthesised by r-process, s-process and rp processes. The drip line 
nuclei play an important role then for the understanding of stellar processes
after the star runs out of fuel. With this in view we have calculated 
the $\beta^\pm$ half lives for the drip line neutron nuclei in the mass 
region $81 < A < 100$.  In this case too the calculations are made by 
constructing bivariate strength densities for $\beta^\pm$ operator under
the framework of spectral averaging theory.

The spectral averaging theory which is based on 
statistical nuclear spectroscopy,
was started with Bethe's statistical mechanical level density formula,
Wigner's treatment of spectral fluctuations using matrix ensembles
and French's embedded ensembles and Gaussian densities.
The smoothed forms of
spectroscopic observables follow from the action of Central Limit
Theorems (CLT) in nuclear shell model spaces. The  
statistical spectroscopy is in deriving and applying the smoothed forms
in indefinitely large spaces with {\it interactions} by using unitary
group decompositions (of Hamiltonians and the spectroscopic spaces),
CLT's locally, and convolutions - the resulting theory is the Spectral
Averaging Theory in Large Shell Model Spaces (SAT-LSS) 
\cite{french,kota1,kota2,kar1,majumdar1,majumdar2}. Here
it is seen that the essential role
of interactions is to produce local spreadings of the non interacting
particle (NIP) densities and the spreadings are in general Gaussian in
nature. The spectral averaging theory in large shell model spaces has 
important nuclear physics applications like calculations of nuclear state 
and level densities, occupation and spin cut-off densities and 
calculations of occupation numbers and spin cut-off factors  \cite{majumdar1}.  

The theory has been extended to calculate the smoothed form for 
interacting particle (IP) bivariate transition strength densities 
({\bf I}$^H_{\cal O}$ $(E_i,E_f)$)
for a one plus two body nuclear Hamiltonian $H$ and a transition operator
${\cal O}$. 
This bivariate strength density  
({\bf I}$^H_{\cal O}$ $(E_i,E_f)$) takes a convolution form 
\cite {kota2,kota3,majumdar3} with the 
non interacting particle (NIP) strength
densities being convoluted with a spreading bivariate Gaussian due to 
irreducible two body part of the interaction. Mathematically, 
{\bf I}$^H_{\cal O} \Rightarrow {\bf I}_{\cal O}^{\bf h} \otimes 
\rho^{\bf V}_{\cal O}$, where {\bf I}$^{\bf h}$ is the strength density 
due to effective one body part {\bf h} of the interacting Hamiltonian 
and $\rho_{\cal O}^{\bf V}$ is a zero centred bivariate Gaussian due 
to the irreducible two body (off-diagonal) part {\bf V} of the interaction. 
In spectral averaging theory the NIP part ${\bf I}_{\cal O}^{\bf h}$  are 
constructed by calculating a few lower order moments of 
${\bf I}_{\cal O}^{\bf h}$ and explicit analytical formulae for NIP strength
densities are worked out in \cite{majumdar3}. This method termed as moment 
method is a convenient way for rapid construction of NIP strength densities.
The IP bivariate strength densities thus constructed has been used to calculate
$\beta^-$ decay rates for some fp-shell nuclei around the mass range 
$55 < A < 65$ \cite{majumdar4}, Giant Dipole Resonance (GDR) cross-sections
\cite{gdr} etc.  

The smoothed form for IP strength densities for $\beta^\pm$ decay and EC 
Gamow-Teller transition operator
can therefore be constructed using the formalism of spectral avearaging 
theory in a large shell model space. With these, the half lives and rates 
for electron capture processes at finite temperature relevant for presupernova
environment and half lives for $\beta^\pm$ decay processes are calculated for 
the nuclei mentioned earlier.

The paper is organised as follows. In Section 2 we give the formalism.
Section 3 deals with the calculational methods that includes choice of 
shell model space, the single particle energies (s.p.e.) etc. The results are
discussed in Section 4. In section 5 we conclude with some discussions and 
future outlook.

\section{Formalism}

\subsection{Unitary decompositon of nuclear Hamiltonian}

Given $m$ number of particles distributed in a shell model space we have an 
$m$ - particle shell model space. Each shell model orbit or single particle
orbit $\alpha$ with 
degeneracy $N_\alpha = 2j_\alpha + 1$ is called a spherical orbit. A group
of spherical orbits is called a unitary orbit {\large \ba}. 
In what follows we will use $\alpha$, $\beta$ etc. (smaller in size) to denote
spherical orbits and {\large \ba}, {\large \bb} etc. (larger in size) 
for unitary orbits. 
Thus we have for $m$
number of particles distributed in this shell model space, spherical 
configuration {\bf m} $\equiv m_\alpha, m_\beta, ...$ ($m_\alpha$, $m_\beta$
etc. are number of particles in spherical orbits $\alpha$, $\beta$ etc. 
respectively)
and unitary configuration [{\bf m}] $\equiv m_{\ba}, m_{\bb},
...$. Using the convention that the spherical orbits $\alpha$ belongs to
unitary orbit {\large \ba} the number of single particle orbits in uniary 
orbit {\large \ba} is $N_{\ba} = \sum_{\alpha \in \ba} N_\alpha$.    

A further decomposition of $m$-particle space is possible by 
attaching an $s_\alpha$ label to each spherical orbit $\alpha$ where    
$s_\alpha$ for lighter nuclei denotes $s\hbar \omega$ excitation value. With
this the $m$- particle space can be decomposed into $S$-subspaces as follows:
\begin{equation}
m \rightarrow \sum S^\pi; S^\pi \rightarrow \sum [{\bf m}]; 
[{\bf m}] \rightarrow \sum {\bf m}; S = \sum m_\alpha s_\alpha
\end{equation}

One can recognise here the appearence of $U(N)$ group which is 
generated by the $N^2$ operators $a^\dagger_{j_\alpha m_\alpha}
a_{j_\alpha m_\alpha}$; $\alpha$, $\beta = 1,2, ..., N$. For $m$ identical
particles therefore there are $\dis \l ( {\begin{tabular}{c} N  \\ 
m \end{tabular}} \r )$ antisymmetric states forming an irreducible
representation (irrep) for the group $U(N)$ usually denoted by Young shape 
$\{ 1^m \}$. The only scalar operator is the number operator $n$ as it remains
invariant under the transformation of the $U(N)$ group. A given operator
can be decomposed into tensor operators (belonging to a definite irrep
of U(N)) with respect to $U(N)$ group. 

Thus the unitary group $U(N_\alpha$) acting on each spherical orbit $\alpha$
generates $m_\alpha$ of spherical configurations {\bf m}, 
i.e. {\bf m} behaves as $\{1^{m_\alpha}\} \otimes \{1^{m_\beta}\} \otimes
....$ with respect to the direct sum group $U(N_\alpha) \oplus U(N_\beta)
\oplus ... $ and  the scalar operators are $m_\alpha$'s. Therefore for 
a given nuclear two body Hamiltonian $ H = h(1) + V(2)$, it should be 
obvious that non interacting particle part $h(1)$ of $H$ is a scalar 
with respect to spherical configuration group and
\be
h = \sum \epsilon_\alpha n_\alpha = h^{[0]}
\ee
where $\epsilon_\alpha$ is the single particle energy (s.p.e.) of the 
spherical orbit $\alpha$. For the spherical configuration scalar part 
$V^{[0]}$ (must be a second order polynomial in $n_\alpha$) of residual 
interaction $V(2)$, one has
\be
V^{[0]} = \dis\sum_{\alpha \ge \beta} V_{\alpha\beta} 
\frac {n_\alpha(n_\beta-\delta_{\alpha\beta})} {(1 + \delta_{\alpha\beta})} 
\,\, .
\ee
In the above $V_{\alpha \beta}$ is the average two body interaction 
given by 
\be
V_{\alpha \beta} = \{ N_{\alpha \beta} \}^{-1} 
\dis \{ \sum_J (2J + 1) V^J_{\alpha\beta\alpha\beta} (1 + \delta_{\alpha\beta})
\}
\ee
where $N_{\alpha \beta} = N_\alpha (N_\beta - \delta_{\alpha\beta})$ 
and $J$ is the angular momentum. The remaining non-diagonal 
part {\bf V} $= V(2) - V^{[0]}$
of $V(2)$ is an irreducible two body part as there cannot be an 
effective one body part of $V(2)$ with respect to spherical configuration
group. 

The same idea can be applied for the decomposition of Hamiltonian under 
unitary configuration group. In this case, the unitary group $U(N_{\ba}$) acting
in each unitary orbit {\large \ba} generates $m_{\ba}$ of unitary configuration 
[{\bf m}]; i.e. [{\bf m}] behaves as $\{1^{m_{\ba}}\} \otimes 
\{1^{m_{\bb}}\} \otimes ....$ with respect to the direct sum 
group $U(N_{\ba}) \oplus U(N_{\bb})
\oplus ... $ and  the scalar operators are $m_{\ba}$'s. The decomposition 
relevant for the present is the spherical scalar part 
$h^{[0]} + V^{[0]}$ of the Hamiltonian. They will have 
$[{\bf 0}] \oplus [{\bf 1}] \oplus [{\bf 2}]$ tensorial parts with respect
to unitary configuration direct sum group. 

Thus after the unitary decomposition of the nuclear Hamiltonian $H$ we
have 
\barr 
H &=& h(1) + V(2)  \nonumber \\ 
& \Rightarrow & h^{[0]} + V^{[0]} + {\bf V} \nonumber \\
&  \Rightarrow & h^{[0][{\bf 0}]} + h^{[0][{\bf 1}]} +V^{[0][{\bf 0}]} +
V^{[0][{\bf 1}]} + V^{[0][{\bf 2}]} + {\bf V}
\earr 

This has been demonstrated that  the contribution of $V^{[0][{\bf 2}]}$ part is
small ($\le 5\%$) all across the periodic table by calculating 
its norm for the case of ds, fp, 10-orbit 
and 15-orbit interactions and phenomenological interactions like
surface delta interaction and pairing + Q.Q interactions \cite{kota1,majumdar1}.
Therefore  $V^{[0][{\bf 2}]}$ can be negleced for all practical purposes. 

Thus Eq. 5 reduces to 
\barr
H&=&  h^{[0][{\bf 0}]} + h^{[0][{\bf 1}]} +V^{[0][{\bf 0}]} +
V^{[0][{\bf 1}]} + {\bf V} \nonumber \\
&=& {\bf h} + {\bf V}
\earr
where {\bf h} is the effective one body part of $H$ and {\bf V} is irreducible
two body part.

For a unitary configuration [{\bf m}], the unitary decompositions of nuclear 
Hamiltonian $H$ are given by
\barr 
h^{[0][{\bf 0}]} &=& \dis\sum_{\ba} \epsilon_{\ba} n_{\ba} \,\, ; \,\,
\epsilon_{\ba} = \l (\dis\sum_{\alpha \in \ba} \epsilon_\alpha N_\alpha \r)
N_{\ba}^{-1} \nonumber \\
&& \nonumber \\
h^{[0][{\bf 1}]} &=& \dis\sum_\alpha \epsilon_\alpha^{[{\bf 1}]}n_\alpha\,\,;
\,\, \epsilon_\alpha^{[{\bf 1}]}= \epsilon_\alpha - \epsilon_{\ba} \nonumber \\
&& \nonumber \\
V^{[0][{\bf 0}]} & = & \dis\sum_{{\ba} \ge {\bb}} [V_{{\ba}{\bb}}]
\frac {n_{\ba}(n_{\bb} - \delta_{{\ba}{\bb}})} 
{(1 + \delta_{\ba\bb})}\,\, ; \,\,
V_{\ba\bb} = 
\dis \l [ \sum_{\alpha \in \ba, \beta \in \bb} N_{\alpha\beta} V_{\alpha\beta}
\r ] \l [ N_{\ba\bb} \r ]^{-1} \nonumber \\
V^{[0][{\bf 1}]} & = & \dis\sum_\alpha 
\l\{\dis\sum_{\bb} (m_{\bb} - \delta_{\ba\bb}) \l 
[ \epsilon_\alpha^{[{\bf 1}];\bb}
\r ] \r \}n_\alpha  \nonumber \\
\epsilon_\alpha^{[{\bf 1}];\bb} &=&
\l \{ \l [ \dis\sum_{\beta \in \bb} (N_\beta - \delta_{\alpha\beta})
V_{\alpha\beta} \r] - (N_{\bb} - \delta_{\ba\bb}) V_{\ba\bb} \r\} \times 
\{ N_{\bb} - 2\delta_{\ba\bb} \}^{-1}
\earr
In the above and for rest of the calculations we consider $S$-conserving 
part (see discussions above Eq. 2 for $S$ quantum number) of the interaction.
The S-mixing part $V_{S-mix}$ of $V(2)$ represents admixing between distant 
configurations (at least 2$\hbar\omega$ away from each other) and this leads
to multimodal form of densities \cite {khm} unlike the unimodal forms. Moreover
GT$\beta^\pm$ operator does not connect different $S$-subspaces.
Hence the $V_{S-mix}$ is not considered for the rest of the calculations. 
  
Given $H = {\bf h} + {\bf V}$ the IP strength density 
${\bf I}^{H={\bf h} + {\bf V}}_{\cal O} (E_i,E_f)$ for a transition 
operator ${\cal O}$ will take a bivariate convolution form \cite{kota3} 
with the two convoluting functions being NIP strength densities
$I^{\bf h}_{\cal O}$ and a normalised spreading bivariate Gaussian 
$ \rho^{\bf V}_{{\cal O};BIV-{\cal G}}$ due to {\bf V} (interactions),   
\begin{equation}
{\bf I}^{H={\bf h} + {\bf V}}_{\cal O} (E_i,E_f) = 
I^{\bf h}_{\cal O} \otimes \rho^{\bf V}_{{\cal O};BIV-{\cal G}} [E_i,E_f]
\end{equation}
For our case, ${\cal O} \equiv {\cal O}(GT)$. In large spectroscopic spaces
with protons and neutrons (pn), this GT bivariate strength density 
can be partitioned in different unitary configuration subspces and 
$S$-subspaces (Eq. 1) and can be written as (identifying that ${\cal O}(GT)$
does not connect two different subspaces) 
$$
{\bf I}^{H={\bf h} + {\bf V}}_{{\cal O}(GT)} (E_i,E_f) = 
\dis\sum_S \sum_{[{\bf m}_p^i,{\bf m}_n^i],[{\bf m}_p^f,{\bf m}_n^f]\in S} 
$$
\be
{\bf I}^{{\bf h};[{\bf m}_p^i,{\bf m}_n^i],[{\bf m}_p^f,{\bf m}_n^f]}_
{{\cal O}(GT)} \otimes 
\rho^{{\bf V};[{\bf m}_p^i,{\bf m}_n^i],[{\bf m}_p^f,{\bf m}_n^f]}
[E_i,E_f]
\ee

In geneal for a transition operator ${\cal O}$ and for a nuclear Hamiltonian 
$H$, the strength density ${\bf I}^{H;(m,m')}$ is given as 
\barr
{\bf I}^{H;(m,m')} &=& I^{m'}(E') | \lan E'm'|{\cal O}| E M\ran|^2 I^m(E) 
\nonumber \\
&=& \lan\lan {\cal O}^\dagger \delta(H - E') {\cal O} \delta (H - E) 
\ran\ran^m
\earr
where $I^{m'}(E')$ and $I^m(E)$ are final and initial state densities 
and $\lan\lan$ $\ran\ran^m$ represents a trace over $m$ particle space.
By the action of CLT in the spectroscopic space, the density will be 
a bivariate Gaussian and it is demonstrated in \cite{majumdar1,majumdar4} 
by constructing the exact NIP strength densities (and its $s$-decomposition)
for ${\cal O}(GT)$ operator and then comparing this with the smoothed 
Gaussian form.   

The smoothed form for strength 
density is constructed with marginal centroids $\epsilon_1$, 
$\epsilon_2$ and variances $\sigma_1^2$, $\sigma_2^2$  
and  calculating few lower order central bivariate
moments given by 
\barr
\epsilon_1 &=& \lan {\cal O}^\dagger {\cal O} H \ran^m / \lan {\cal O}^\dagger 
{\cal O} \ran^m \nonumber \\
\epsilon_2 &=& \lan {\cal O}^\dagger H {\cal O} \ran^m / \lan {\cal O}^\dagger 
{\cal O} \ran^m \nonumber \\
\sigma_1^2 &=& \lan {\cal O}^\dagger {\cal O} H^2 \ran^m / 
\lan {\cal O}^\dagger {\cal O} \ran^m \nonumber \\
\sigma_2^2 &=& \lan {\cal O}^\dagger H^2 {\cal O} \ran^m / 
\lan {\cal O}^\dagger {\cal O} \ran^m \nonumber \\
\mu_{pq} &=& \dis  \l \lan {\cal O}^\dagger \l ( \frac {H - \epsilon_2} {\sigma_2}
\r )^q {\cal O}\l ( \frac {H - \epsilon_1} {\sigma_1} \r )^p  \r \ran^m 
{\Big /} \lan {\cal O}^\dagger {\cal O} \ran^m
\earr 
Therefore the calculation of moments are in fact calculation of $m$-particle
averages or traces of the operator in question. This is done by first 
calculating a few basic traces and then propagating them in m-particle space.
As here we are dealing with unitary configurations [{\bf m}] and unitary 
configuration densities (Eq. 9), we would require the unitary configuration 
traces of the type $\lan$ $\ran^{[{\bf m}]}$. These moments $M_{pq}([{\bf m}])$
for the construction of NIP strength densities {\bf I}$^{\bf h}_{\cal O}$
for a one body transition operator ${\cal O}$ with $p+q \leq 2$ are calculated 
in details in Ref.  \cite{majumdar1,majumdar4}.   As the GT operator is 
of the type ${\cal O}(GT) = \epsilon_{\alpha\beta}a^\dagger_\alpha a_\beta$
($\epsilon_{\alpha\beta}$ is the single particle matrix elements), for a 
given initial configuration [{\bf m}$_i$], the final configuration 
[{\bf m}$_f$] then is obtained uniquely
as $[{\bf m}_f] = [{\bf m}_i] \times \l ( 1_\alpha^\dagger 1_\beta \r )$. Thus 
one also obtains the partial moments $M_{pq}([{\bf m}_i],[{\bf m}_f])$ for 
{\bf I}$^{\bf h}$. For a proton neutron (pn) configuration one writes [{\bf m}]
as [{\bf m}$_p$,{\bf m}$_n$]. 

For the construction Gaussian spreadings 
$\rho^{{\bf V};[{\bf m}_p^i,{\bf m}_n^i];[{\bf m}_p^f,{\bf m}_n^f]}_
{\cal O}(GT) (x,y)$, in Eq. 9, the following approximations are adopted. 
The marginal centroids, $M_{10} = \lan {\cal O}^\dagger {\cal O} {\bf V}
\ran^{m_p^i,m_n^i} / \lan {\cal O}^\dagger {\cal O} \ran^{m_p^i,m_n^i}$ 
$\simeq$ $\lan {\bf V} \ran^{m_p^i,m_n^i} = 0$, as {\bf V} is traceless; 
$M_{01} = \lan {\cal O}^\dagger {\bf V} {\cal O} 
\ran^{m_p^f,m_n^f} / \lan {\cal O}^\dagger {\cal O} \ran^{m_p^f,m_n^f}$ 
$\simeq$ $\lan {\bf V} \ran^{m_p^f,m_n^f} = 0$ and the marginal variances
given by traces of the type $\lan {\cal O}^\dagger {\cal O} {\bf V}^2\ran$
and $\lan {\cal O}^\dagger {\bf V}^2 {\cal O} \ran$ are equal to 
$\lan {\bf V}^2 \ran$. 

\subsection{Formalism for $\beta^\pm$ decay half lives and electron
capture rates at finite temperature}

The weak interaction ($\beta^\pm$ and electron capture (EC) for the 
present case) rate $T(E_i \rightarrow E_f)$ is the number of 
weak processes per second from a given initial state $|E_i\rangle$ 
of the parent 
nucleus to the final nuclear state $|E_f\rangle$ and 
$T (E_i \rightarrow E_f) \propto [g_V^2 B_F (E_i \rightarrow E_f) +
g_A^2 B_{GT}(E_i \rightarrow E_f)]$, where $g_V$ and $g_A$ are respectively
the vector and axial vector coupling constants and $B_F$ and $B_{GT}$ are 
the Fermi and Gamow - Teller transition strengths respectively. Including 
the phase space factor $f$ that incorporates the dependence of the rate on 
nuclear charge $Ze$ and the available energy for the weak process 
under consideration, $T$ takes 
the form $T (E_i \rightarrow E_f) = Cf[g_V^2B_F
(E_i \rightarrow E_f) + g_A^2 B_{GT} (E_i \rightarrow E_f)]$ where $C$ is 
a constant with $Q$ the GS $Q$-value of the weak process ($\beta^\pm$ or EC),
$Q_i = Q + E_i$. The value of $C$ is fixed by using the $\ell og ft$ values 
from pure Fermi transitions. One can write down the expressions for ground 
state half lives and $\beta^\pm$ decay and EC rates at finite temperature. 
For present beta decay calculations we have neglected  
Fermi term $B_F$ as the Fermi strength is concentrated in narrow 
domain and high up in energy (the centroid $\sim 1.44Z A^{-1/3}$ MeV, width
$\sim 0.157Z A^{-1/3}$ MeV). For $\beta^+$ decay and electron capture 
processes, Fermi transitions are not possible for nuclei with $N > Z$, 
since for mother isospin $T_0 (=(N-Z)/2)$, the isospins 
of the states of daughter are greater by 1, {\em i.e.}
$T_{daughter} = \left |\{ (N+1) - (Z-1) \}/2 \right | = T_0 + 1$. 
Writing $B_{GT}(J_i E_i \rightarrow J_f E_f) 
= (2J_i + 1)^{-1} \sum |\langle E_f J_f M_f | ({\cal O}_{GT})^k_\mu | E_iJ_iM_i
\rangle^2$ which in continuous version becomes $\{I(E_i)\}^{-1} \sum_{\alpha
\in E_i, \beta \in E_f , \mu} |\langle E_f \alpha |
({\cal O}_{GT})^k_\mu | E_i \beta \rangle |^2$, the expression for 
GS half life is 
\begin{center}
$t_{1/2}(GS) = \{ 6250 (s) \} \times$ 
\end{center}
\begin{equation}
\left\{ \displaystyle\int^Q_0 \left [ \left ( \frac {g_A} {g_V} \right)^2 
3 \pounds \right ] \left [ \frac {{\bf I}^H_{{\cal O}(GT)} (E_{GS}, E_f)}
{I^H(E_{GS})} \right ] f(Z) dE_f \right \}^{-1}
\end{equation}
In the above equation $\pounds$ is the so called quenching factor and 
the factor 
3 comes because of the definition of {\bf I}$^h$. The usual values of $\pounds$
is 0.6 \cite{ho83} for $\beta^-$ decay and 0.5 for EC/$\beta^+$ decay 
and $\displaystyle\left(\frac {g_A} {g_V} \right)^2 = 1.4$ 
as given in \cite{br77}. The weak interaction rates are given by
\barr
\lambda(T) &=& \frac {{\ell n}2(s^{-1})} {6250} 
\l [ \int e^{-E_i/k_BT} I^H (E_i) dE_i \r ]^{-1} \times \nonumber \\ 
&&\l [ \int dE_i e^{-E_i/k_BT} I^H(E_i) \l [ \int^{Q_i}_0 dE_f \l\{ 
\l ( \frac {g_A} {g_V} \r )^2 3\pounds \r\} \times \r. \r. \nonumber \\
&& \l.\l.  \l [ \frac {I^H_{{\cal O}(GT)} (E_i, E_f)} {I^H (E_i)} \r ]
f(Z,T) \r ] \r ] \nonumber \\
&& \nonumber \\
&=& \frac {{\ell n}2(s^{-1})} {6250}
\l [ \int e^{-E_i/k_BT} I^H (E_i) dE_i \r ]^{-1} \times \nonumber \\
&&\l [ \int dE_i e^{-E_i/k_BT} \l [ \int^{Q_i}_0 dE_f \l\{
\l ( \frac {g_A} {g_V} \r )^2 3\pounds \r\} 
I^H_{{\cal O}(GT)} (E_i, E_f)
f(Z,T) \r ] \r ] \nonumber \\
&&
\earr 
The phase space integral $f$ in the above eqautions are given by, for 
the case of EC \cite{murphy} 
\barr
f &=& \int^{\infty}_{E_{min}} dE E (E^2 - m_e^2)^{1/2} (E - Q_i)^2 F_c (Z,E)
\times \nonumber \\
&& \,\,\,\, \l (\frac {1} {1 + exp [(E - \mu_e)/k_B T]}\r )
\l ( 1 - \frac {1} {1 + exp [(E - \mu_nu)/k_B T]} \r )
\earr
where $E_{min}$ is greater among $-Q_i$ and $m_ec^2$. 
In the actual calculation the integrand is expressed in terms of electron 
mass $m_e$ and the integration limits are also accordingly changed. 
In the above $k_B$ stands for Boltzmann constant, $\mu_e$ is the chemical 
potential for electron and $\mu_\nu$ is the same for neutrino. 
For the case of $\beta^-$ decay the phase space factor reads as
\barr
f &=& \int^{\epsilon_0}_1 \frac {F_c(Z,\epsilon_e) \epsilon_e 
(\epsilon_e^2 -1)^{1/2} (\epsilon_0 - \epsilon_e )^2} 
{ \{1 + exp[(\mu_e - \epsilon_e)/(k_B T)_e ] \} } d\epsilon_e 
\earr
In the above equation $\epsilon_0 = E_0/m_e$; ($E_0 = Q_i - E_f$), 
$\mu_e = \mu/m_e$ and 
$(k_B T)_e = k_B T/ m_e$. $F_c(Z,E)$ in the above equations give the Coulomb
factor and for EC is given by \cite{murphy}
\barr
F_c(Z,E) &=& F_c^\prime (Z) F_e^\prime (E) \nonumber \\ 
F_c^\prime &=& \frac {8\pi\alpha Z (1 + S)} {|\Gamma (1+2s) |^2} 
\l ( \frac {2R} {\hbar} \r )^{2S-2} \l(\frac {1 + 0.577 (\alpha Z)^2} 
{2 + 0.577 (\alpha Z)^2}\r) \nonumber \\
F_e^\prime (E) &=& (E^2 - m_e^2)^{S-1} \l(\frac {E} {(E^2 - m_e^2)^{1/2}}\r) 
\l(\frac {1} {1 - e^{-2\pi\theta}}\r)
\earr
In the above, $\alpha$ is the fine structure constant. The value of $S$ 
in the present calculation is assumed to be 1 \cite{murphy} and following
\cite{murphy} the term $\l(\frac {1} {1 - e^{-2\pi\theta}}\r)$ is expressed as
$\lan 2\pi \theta \ran$  and for $Q_i < m_e c^2$ 
\barr
\lan 2\pi \theta \ran &=& 2\pi\alpha Z \frac {exp (-1/\lambda)} 
{K_2 (1/\lambda)} (1 + 2\lambda + 2\lambda^2) \\
\lambda &=& k_B T/m_e c^2 \nonumber
\earr
where $K_2$ is second order modified Bessel function. For $Q_i > m_e c^2$
we use
\barr
\lan 2\pi \theta \ran &=& 2\pi\alpha Z \frac {Q_i} {(\epsilon_0^2 - m_e^2)^{1/2}}
\earr

For $\beta^-$ process the Coulomb factor is taken as given in Schenter and Vogel
\cite{sc83} and the expression reads as
\barr
F_c(Z,\epsilon) &=& \frac {\epsilon} {\sqrt {\epsilon^2 -1}} exp [\alpha(Z) 
+ \beta(Z) \sqrt {\epsilon -1}]; \nonumber \\
\alpha(Z) &=& -0.811 + 4.46(-2)Z + 1.08(-4)Z^2 \,\,\,\,\,\,\,\,\,\,\,\,\,\,\,
\,\,\,\,\,\,\,\, (\epsilon -1)<1.2
\nonumber \\
&=& -8.46(-2) + 2.48(-2)Z + 2.37(-4)Z^2 \,\,\,\,\,\,\,\,\,\,\,  
(\epsilon -1)\ge 1.2 
\nonumber \\
\beta(Z) &=& 0.673 - 1.82(-2)Z + 6.38(-5)Z^2  \,\,\,\,\,\,\,\,\,\,\,\,\,\,\,
\,\,\,\,\,\,\,\,\, (\epsilon -1)<1.2
\nonumber \\
&=& 1.15(-2) + 3.58(-4)Z - 6.17(-5)Z^2 \,\,\,\,\,\,\,\,\,\,\,\,\,\,\,
(\epsilon -1)\ge 1.2
\nonumber \\
&& 
\earr       
 
For the electron chemical potential we have used theexpression given in 
\cite {au},
\barr 
\mu_e &=& 1.11 (\rho_7 Y_e)^{1/3} \l [ 1 + \l ( \frac {\pi} {1.11} \r)^2 
\frac {{\bar T}^2} {(\rho_7 Y_e)^{2/3}} \r]^{-1/3}.
\earr
In the above $\rho_7$ is the matter density $\rho$ in units of 10$^7$ gms/cc,
${\bar T}$ is the temperature $T$ expressed in MeV and $Y_e$ is the 
electron fraction. We put $\mu_\nu$, the neutrino chemical potential to be 
zero as for the densities we consider the neutrinos are free streaming.

\section{Calculational Procedure}

For the construction of strength densities we have selected a 
9-orbit shell model space both for proton and neutron with $^{56}Ni$ as
core, consisting of the spherical orbits
$3p_{3/2}$, $3f_{5/2}$, $3p_{1/2}$, $4g_{9/2}$, $4d_{5/2}$, $4g_{7/2}$,
$4s_{1/2}$, $4d_{3/2}$ and $5h_{11/2}$. The s.p.e. in MeV are 
26.58, 26.19, 29.09, 33.91, 38.52, 42.47, 42.30, 43.15, 58.44 respectively.
The initial values of s.p.e's are taken from ref \cite{hillman}. 
The renormalisation effects 
due to the closed core (in this case $s$ shell, $p$ shell, $ds$ shell 
and $f_{7/2}$ orbit) are then incorporated. This effect not only renormalises
$fp - sdg$ and $sdg - h_{11/2}$ separation but also renormalises the single
particle energies. They can be evaluated from $V_{\alpha\beta}$ and 
$V_{\ba\bb}$ discussed in Sect. 2.1. 
The unitary orbits are \{$3p_{3/2}$, $3f_{5/2}$, $3p_{1/2}$\},
\{$4g_{9/2}$\}, \{$4d_{5/2}$, $4g_{7/2}$, $4s_{1/2}$,$4d_{3/2}$\},
\{$5h_{11/2}$\}. Thus each of the proton and neutron shell model space
has been divided into four unitary orbits.  
For the 2-body residual interaction, we have used a phenomenological
interaction namely pairing + Q.Q interaction \cite{baranger} with 
the strength $\chi = 242/A^{-5/3}$.  

The nuclei chosen for different rate and half life calculations are given
in Table 1. These nuclei have been chosen from Ref. \cite{schatz}

Using the formalism given in Sect. 2 one can now construct the weak interaction
strength densities (bivariate Gaussian form). For the state densities 
$I^H(E)$, required for calculation of half lives or rates (Eq. 12, 13), we 
adopted the formula given in Dilg et al \cite{dilg}. This has been demonstrated 
in \cite{majumdar1,majumdar2} that state densities $I^H(E)$ obtained from 
spectral averaging theory represent very well the results obtained 
from Dilg et al state density formula.   
Besides the assumption of 
marginal centroids and variances (Sect. 2.1), we also assume the bivariate
correlation coefficient $\zeta$ to be independent of configuration. Moreover,
for two-body EGOE (Embedded Gaussian Orthogonal Ensemble) \cite{kota2}, 
one has the result $\zeta \simeq 1 - 2/m$, where $m$ is the number of 
active nucleons. Therefore, in the present calculation, $\zeta$ is first 
evaluated by treating it as a parameter with the EGOE form for 
$\zeta = a + b/m$ 
and $a$ and $b$ is evaluated by calculating $\beta^-$ decay half lives 
for 18 nuclei in the mass range considered and
then comparing them with the experimental values of the same. 

To this end, we fix $a$ and $b$ by calculating $\zeta$ 
for various values of $a$ and $b$ with the constraint that the value of 
$\zeta$ lies within 0.6 and 0.9. The value of bivariate variance (approximated
as state density variance, $\lan {\bf V}^2 \ran$ (Sect. 2.1)) is
taken to be 16.5 MeV$^2$. The values of $a$ and $b$ are found out 
by minimizing the quantity $\displaystyle\sum_{i=nuclei} 
(\ell og(\tau^i_{1/2})_{cal} - \ell og(\tau^i_{1/2})_{expt})^2$. The values of 
$a$ and $b$ are obtained respectively as 0.77 and 0.3. 
The values of $\zeta$ and $\beta^-$ half
lives for the nuclei considered are listed in the Table 2. 

Having obtained the values of the parameters $a$ and $b$ for calculation 
of $\zeta$, we now proceed to calculate $\beta^+$ decay half lives for the 
nuclei given in Table 1. For this calculation, we keep $\zeta$ unchanged
from the value obtained for $\beta^-$ decay 
half life calculations. But here we have parametrised the  
variance as $\sigma^2_{\bf V} = A + B/m$ and obtained the parameters 
$A$ and $B$ by minimising  
$\displaystyle\sum_{i=nuclei} 
(\ell og(\tau^i_{1/2})_{cal} - \ell og(\tau^i_{1/2})_{expt})^2$ for the 
four nuclei considered for $\beta^+$ decay (Table 1) much the same 
way as determination of $\zeta$ in $\beta^-$ decay calculations. 
For calculating $\beta^+$ half lives
we use  Eqs. 12 and Eqs 16 - 18. The values of $A$ and $B$ are 
obtained as $A = 20.5$ and $B = 1.9$. The values of calculated half lives
and the variance $\sigma^2_{\bf V}$ for the four nuclei considered for 
$\beta^+$ decay are shown in Table 3.

Calculation of electron capture (EC) half lives are similar to that 
of $\beta^+$ calculations. We use the parameters $a$ and $b$ obtained 
from $\beta^-$ calculations to obtain correlation coefficients $\zeta$
for the five nuclei (Table 1) chosen for EC calculations. The variance
$\sigma^2_{\bf V}$ is parametrized as $\sigma^2_{\bf V} = A + B/m$ 
and the values of $A$, $B$ and consequently $\sigma^2_{\bf V}$ are
found out by minimising $\displaystyle\sum_{i=nuclei}
(\ell og(\tau^i_{1/2})_{cal} - \ell og(\tau^i_{1/2})_{expt})^2$. 
In this case the values of $A$ and $B$ are obtained as 
$A = 11.5$, $B = -4.6$.
For these calculations 
we have used the Eqs. 12, 14, 16 - 18. The calculated half lives and 
$\sigma^2_{\bf V}$ are shown in Table 4.

We have also calculated EC decay rates for typical densities and 
temperatures relevant for presupernova stars. 
The rates are evaluated for three different temperatures ($3 \times 
10^9$, $4 \times 10^9$ and $5 \times 10^9$ in $^oK$), three different 
stellar densities ($10^9$, $10^8$ and $10^7$ gms/cc) and for three 
values for electron fraction $Y_e$ (0.50, 0.47, 0.43). 
The calculations use Eqs. 13,14, 15-18 and Eq. 20 are used. 
The calculated EC rates alongwith temperatures,
$Y_e$ and stellar densities for each of the 
five nuclei considered, are given in Table 5.

\section{Discussions and Conclusions}

We have calculated the half lives and rates for weak interactions 
for a number of nuclei spanning a wide range ($65 < A < 100$) 
in the periodic table. These calculations are performed by explicitly 
constructing  smoothed form for bivariate strength densities 
for the weak interaction operator
(in this case Gamow Teller operator) using convolution form within 
the framework of spectral averaging theory in nuclear physics.
The lower mass region of the above range 
is relevant for presupernova and supernova stars and the higer mass
region is relevant for r-process nucleosynthesis that synthesises 
heavier nuclei. 
We have used here, the principles of spectral averaging theory 
for calculation of EC rates for different temperatures, densities and
electron fraction values in presupernova environment. Earlier such calculations
with spectral averaging theory have been performed for $\beta^-$ 
decay rates for presupernova stars and for a few fp-shell nuclei \cite{kas}. 
The $\beta^+$ and $\beta^-$ decay half lives are calculated for the nuclei in 
the range $A > 80$. The calculation of the rates for typical r-process 
conditions  for these processes will be taken up in near future.  

\vskip 1cm      
The authors thank A. Ray for helpful discussions.

\newpage
\begin{center}
{\bf Table Captions}
\end{center}

\noindent {\bf Table 1} Nuclei chosen for different weak processes.

\noindent {\bf Table 2} Calculated and experimental 
$\beta^-$ decay half lives and correlation coefficient $\zeta$. Method
of calculating $\zeta$ is given in the text.  

\noindent {\bf Table 3} Calculated and experimental 
$\beta^+$ decay half lives and variance $\sigma_{\bf V}^2$. $Q$-values are
also given. 

\noindent {\bf Table 4}  Calculated and experimental 
EC half lives and variance $\sigma_{\bf V}^2$. $Q$-values are
also given.

\noindent {\bf Table 5} Electron Capture rates for 
$^{65}$Ge, $^{69}$Se, $^{73}$Kr, $^{77}$Sr, $^{81}$Zr. 

\newpage

\begin{center}
{\bf Table 1}
\end{center}

\begin{center}
\begin{tabular}{|c|l|}
\hline
Weak Interaction & Nuclei Chosen \\
\hline
Electron Capture &  $^{65}_{32}Ge_{33}$, $^{69}_{34}Se_{35}$, 
$^{73}_{36}Kr_{37}$, \\
&  $^{77}_{38}Sr_{39}$,$^{81}_{40}Zr_{41}$ \\
\hline
$\beta^+$ decay &  $^{85}_{42}Mo_{43}$, $^{89}_{44}Ru_{45}$, 
$^{93}_{46}Pd_{47}$, \\
&$^{97}_{48}Cd_{49}$ \\
\hline
$\beta^-$ decay & $^{81}_{30}Zn_{51}$, $^{81}_{31}Ga_{50}$, 
$^{81}_{33}As_{48}$ \\
& $^{82}_{32}Ge_{50}$,  $^{83}_{31}Ga_{52}$,  $^{83}_{33}As_{50}$ \\
& $^{84}_{32}Ge_{52}$,  $^{85}_{34}Se_{51}$,  $^{87}_{33}As_{54}$ \\
& $^{88}_{37}Rb_{51}$,  $^{89}_{34}Se_{55}$,  $^{89}_{35}Br_{54}$ \\
& $^{90}_{36}Kr_{54}$,  $^{91}_{37}Rb_{54}$,  $^{92}_{35}Br_{57}$ \\
& $^{92}_{38}Sr_{54}$,  $^{93}_{36}Kr_{57}$,  $^{94}_{39}Y_{55}$ \\
\hline
\end{tabular}
\end{center}

\newpage
\begin{center}
{\bf Table 2}
\end{center}
\begin{center}
\begin{tabular}{|c|c|c|c|c|c|c|}
\hline
Sr No. & Nucl. & $Z$ & $Q_{val}$  & $T_{1/2}^{expt.}$ & $T_{1/2}^{calc}$ & $\zeta$ \\
   &   &     &  (MeV)     &                   &                  & \\
\hline
1 & $^{81}Zn$ & 30 & 11.9 & 0.29 & 0.819 & 0.7820 \\
\hline
2 & $^{81}Ga$ & 31 & 8.32 & 1.221&  1.483 & 0.7820 \\
\hline
3 & $^{81}As$ & 33 & 3.86 & 33.3&  3.662 & 0.7820 \\
\hline
4 & $^{82}Ge$ & 32 & 4.7 & 4.6 & 53.088 & 0.7815 \\
\hline
5 & $^{83}Ga$ & 31 & 11.5 & 0.31&  0.127&  0.7811 \\
\hline
6 & $^{83}As$ & 33 & 5.46 & 13.4 & 21.134 & 0.7811 \\
\hline
7 & $^{84}Ge$ & 32 & 7.7 & 1.2 & 2.415 & 0.7807 \\
\hline
8 & $^{85}Se$ &  34 & 6.182 & 31.7&  106.261&  0.7803 \\
\hline
9 & $^{87}As$ &  33 & 10.3 & 0.73 & 0.0632 & 0.7797 \\
\hline
10 & $^{88}Rb$ &  37 & 5.13&  1066.8 & 27227.014 & 0.7794 \\
\hline
11 & $^{89}Se$ &  34 & 9.0 & 0.41 & 0.117 & 0.7791 \\
\hline
12 & $^{89}Br$ &  35 & 8.16&  4.4&  1.465 & 0.7791 \\
\hline
13 & $^{90}Kr$ &  36&  4.39&  32.32&  413.661&  0.7788 \\
\hline
14 & $^{91}Rb$ &  37&  5.86 & 58.4 & 290.453&  0.7786\\
\hline
15 & $^{92}Br$ & 35&  12.2 & 0.343 & 0.096 & 0.7783 \\
\hline
16 & $^{92}Sr$ &  38&  1.67&  9756.&  110696.591 & 0.7783 \\
\hline
17 & $^{93}Kr$ &  36&  8.6 & 1.286 & 3.706 & 0.7781 \\
\hline
18 & $^{94}Y$  & 39&  4.0 & 67320. & 2590.043 & 0.7779 \\
\hline
\end{tabular}
\end{center}

\newpage
\begin{center}
{\bf Table 3}
\end{center}
\begin{center}
\begin{tabular}{|c|c|c|c|c|c|}
\hline
Nucleus & Z & $Q_{\beta^+}$ & $T_{1/2}^{exp}$ & $T_{1/2}^{Calc}$ & 
$\sigma^2_{\bf V}$  \\
        &   & (MeV)    &  sec          &   sec            &  MeV$^2$    \\
\hline
$^{85}$Mo & 42&  8.6 & 0.4 &  0.076 &  20.566 \\
\hline
$^{89}$Ru & 44&  8.92&  0.3&   0.22 &  20.558  \\
\hline
$^{93}$Pd & 46&  9.5 & 0.3 &  0.14 & 20.551   \\
\hline
$^{97}$Cd & 48&  9.6 & 0.2 &  0.32 & 20.546     \\   
\hline
\end{tabular}
\end{center}

\vskip 2cm

\begin{center}
{\bf Table 4}
\end{center}
\begin{center}
\begin{tabular}{|c|c|c|c|c|c|}
\hline
Nucleus & Z & $Q_{EC}$ & $T_{1/2}^{exp}$ & $T_{1/2}^{Calc}$ & $\sigma^2$  \\
        &   & (MeV)    &  sec          &   sec            &  MeV$^2$    \\
\hline
$^{65}$Ge &  32. & 6.24 & 30.9 &  12.79 & 10.989 \\
\hline
$^{69}$Se &  34. & 6.78 & 27.4 &  10.32 & 11.146 \\
\hline
$^{73}$Kr &  36. & 6.65 & 27.  & 12.58 & 11.229  \\
\hline
$^{77}$Sr &  38. & 6.85 & 9.   &  9.35 & 11.281  \\
\hline
$^{81}$Zr &  40. & 7.16 & 15.  & 14.04 & 11.316     \\
\hline
\end{tabular}
\end{center}

\newpage

\begin{center}
{\bf Table 5}
\end{center}

\begin{center}
\begin{tabular}{|c|c|c|c|c|c|}
\hline
Nucleus & $\rho$(gms/cc) &  & \multicolumn{3}{|c|} {Temperature in $^o$K} \\
\cline{4-6}
& & $Y_e$ & $3 \times 10^9$ & $4 \times 10^9$ & $5 \times 10^9$ \\
\cline{4-6}
& &       & \multicolumn{3}{|c|}{Rates ($s^{-1}$)} \\
\hline
$^{65}$Ge & $10^9$ & 0.50 & $2.77 \times 10^{-3}$ & $2.89 \times 10^{-3}$
& $3.07 \times 10^{-3}$ \\
 &  & 0.47 & $2.56 \times 10^{-3}$ & $2.68 \times 10^{-3}$
& $2.86 \times 10^{-3}$ \\
 &  & 0.43 & $2.28 \times 10^{-3}$ & $2.41 \times 10^{-3}$
& $2.58 \times 10^{-3}$ \\
& $10^8$ & 0.50 & $2.05 \times 10^{-4}$ & $2.47 \times 10^{-4}$
& $3.13 \times 10^{-4}$ \\
&  & 0.47 & $1.92 \times 10^{-4}$ & $2.34 \times 10^{-4}$
& $2.98 \times 10^{-4}$ \\
&  & 0.43 & $1.76 \times 10^{-4}$ & $2.16 \times 10^{-4}$
& $2.79 \times 10^{-4}$ \\
& $10^7$ & 0.50 & $2.96 \times 10^{-5}$ & $5.67 \times 10^{-5}$
& $1.10 \times 10^{-4}$ \\
&  & 0.47 & $2.86 \times 10^{-5}$ & $5.56 \times 10^{-5}$
& $1.10 \times 10^{-4}$ \\
&  & 0.43 & $2.71 \times 10^{-5}$ & $5.43 \times 10^{-5}$
& $1.09 \times 10^{-4}$ \\
\hline
$^{69}$Se & $10^9$ & 0.50 & $2.81 \times 10^{-3}$ & $2.92 \times 10^{-3}$
& $3.10 \times 10^{-3}$ \\
& & 0.47 & $2.60 \times 10^{-3}$ & $2.72 \times 10^{-3}$
& $2.89 \times 10^{-3}$ \\
& & 0.43 & $2.32 \times 10^{-3}$ & $2.45 \times 10^{-3}$
& $2.62 \times 10^{-3}$ \\
& $10^8$ & 0.50 & $2.20 \times 10^{-4}$ & $2.64 \times 10^{-4}$
& $3.31 \times 10^{-4}$ \\
& & 0.47 & $2.07 \times 10^{-4}$ & $2.50 \times 10^{-4}$
& $3.16 \times 10^{-4}$ \\
& & 0.43 & $1.89 \times 10^{-4}$ & $2.31 \times 10^{-4}$
& $2.96 \times 10^{-4}$ \\
& $10^7$ & 0.50 & $3.26 \times 10^{-5}$ & $6.16 \times 10^{-5}$
& $1.19 \times 10^{-4}$ \\
& & 0.47 & $3.14 \times 10^{-5}$ & $6.05 \times 10^{-5}$
& $1.18 \times 10^{-4}$ \\
& & 0.43 & $2.99 \times 10^{-5}$ & $5.90 \times 10^{-5}$
& $1.17 \times 10^{-4}$ \\
\hline
$^{73}$Kr & $10^9$ & 0.50 & $2.18 \times 10^{-3}$ & $2.27 \times 10^{-3}$
& $2.40 \times 10^{-3}$ \\
& & 0.47 & $2.01 \times 10^{-3}$ & $2.11 \times 10^{-3}$
& $2.24 \times 10^{-3}$ \\
& & 0.43 & $1.80 \times 10^{-3}$ & $1.90 \times 10^{-3}$
& $2.03 \times 10^{-3}$ \\
& $10^8$ & 0.50 & $1.72 \times 10^{-4}$ & $2.05 \times 10^{-4}$
& $2.58 \times 10^{-4}$ \\
& & 0.47 & $1.61 \times 10^{-4}$ & $1.95 \times 10^{-4}$
& $2.46 \times 10^{-4}$ \\
& & 0.43 & $1.48 \times 10^{-4}$ & $1.80 \times 10^{-4}$
& $2.30 \times 10^{-4}$ \\
& $10^7$ & 0.50 & $2.55 \times 10^{-5}$ & $4.82 \times 10^{-5}$
& $9.24 \times 10^{-5}$ \\
& & 0.47 & $2.46 \times 10^{-5}$ & $4.73 \times 10^{-5}$
& $9.19 \times 10^{-5}$ \\
& & 0.43 & $2.34 \times 10^{-5}$ & $4.61 \times 10^{-5}$
& $9.13 \times 10^{-5}$ \\
\hline
\end{tabular}
\end{center}

\newpage
\noindent {\bf Table 5.} Contd.
\begin{center}
\begin{tabular}{|c|c|c|c|c|c|}
\hline
Nucleus & $\rho$(gms/cc) &  & \multicolumn{3}{|c|} {Temperature in $^o$K} \\
\cline{4-6}
& & $Y_e$ & $3 \times 10^9$ & $4 \times 10^9$ & $5 \times 10^9$ \\
\cline{4-6}
& &       & \multicolumn{3}{|c|}{Rates ($s^{-1}$)} \\
\hline
$^{77}$Sr & $10^9$ & 0.50 & $1.86 \times 10^{-3}$ & $1.90 \times 10^{-3}$
& $1.98 \times 10^{-3}$ \\
& & 0.47 & $1.72 \times 10^{-3}$ & $1.77 \times 10^{-3}$
& $1.86 \times 10^{-3}$ \\
& & 0.43 & $1.54 \times 10^{-3}$ & $1.60 \times 10^{-3}$
& $1.68 \times 10^{-3}$ \\
& $10^8$ & 0.50 & $1.56 \times 10^{-4}$ & $1.84 \times 10^{-4}$
& $2.26 \times 10^{-4}$ \\
& & 0.47 & $1.47 \times 10^{-4}$ & $1.74 \times 10^{-4}$
& $2.16 \times 10^{-4}$ \\
& & 0.43 & $1.35 \times 10^{-4}$ & $1.62 \times 10^{-4}$
& $2.03 \times 10^{-4}$ \\
& $10^7$ & 0.50 & $2.39 \times 10^{-5}$ & $4.41 \times 10^{-5}$
& $8.27 \times 10^{-5}$ \\
& & 0.47 & $2.30 \times 10^{-5}$ & $4.33 \times 10^{-5}$
& $8.23 \times 10^{-5}$ \\
& & 0.43 & $2.19 \times 10^{-5}$ & $4.22 \times 10^{-5}$
& $8.18 \times 10^{-5}$ \\
\hline
$^{81}$Zr & $10^9$ & 0.50 & $1.45 \times 10^{-3}$ & $1.48 \times 10^{-3}$
& $1.55 \times 10^{-3}$ \\
& & 0.47 & $1.34 \times 10^{-3}$ & $1.38 \times 10^{-3}$
& $1.44 \times 10^{-3}$ \\
& & 0.43 & $1.20 \times 10^{-3}$ & $1.24 \times 10^{-3}$
& $1.31 \times 10^{-3}$ \\
& $10^8$ & 0.50 & $1.15 \times 10^{-4}$ & $1.36 \times 10^{-4}$
& $1.68 \times 10^{-4}$ \\
& & 0.47 & $1.09 \times 10^{-4}$ & $1.29 \times 10^{-4}$
& $1.60 \times 10^{-4}$ \\
& & 0.43 & $9.96 \times 10^{-5}$ & $1.19 \times 10^{-4}$
& $1.50 \times 10^{-4}$ \\
& $10^7$ & 0.50 & $1.72 \times 10^{-5}$ & $3.20 \times 10^{-5}$
& $6.04 \times 10^{-5}$ \\
& & 0.47 & $1.66 \times 10^{-5}$ & $3.14 \times 10^{-5}$
& $6.01 \times 10^{-5}$ \\
& & 0.43 & $1.58 \times 10^{-5}$ & $3.06 \times 10^{-5}$
& $5.97 \times 10^{-5}$ \\
\hline
\end{tabular}
\end{center}


\begin{thebibliography}{99}
\bibitem{ffn} G.M. Fuller, W.A. Fowler and M.J. Newman, Ap. J. Suppl. 
{\bf 42}, 447 (1980); Ap. J. {\bf 252}, 715 (1982); Ap. J. Suppl. {\bf 48},
279 (1982); Ap. J. {\bf 293}, 1 (1985).
\bibitem{au} M.B. Aufderhide, G.E. Brown, T.T.S. Kuo, D.B. Stout and P. Vogel,
Ap. J {\bf 362}, 241 (1990).
\bibitem{au2} M.B. Aufderhide, I Fushiki, S.E. Woosely and D.H. Hartmann, 
Ap. J. Supp. {\bf 91}, 389 (1994).
\bibitem{kar0} K. Kar, S.Sarkar and A. Ray, Phys. Lett. B {\bf 261}, 217 (1991);
Ap. J {\bf 434}, 662 (1994).
\bibitem{majumdar4} V.K.B. Kota and D. Majumdar, Z. Phys. A {\bf 351}, 
377 (1995).
\bibitem{french} J.B. French, {\it Nuclear Structure}, ed. A. Hossain,
Harun-ar-Rashid and M. Islam (North Holland, Amsterdam)
\bibitem{kota1} J.B. French and V.K.B. Kota, Phys. Rev. Lett. {\bf 51}, 2183
(1983); 
{\it University of Rochester Report}, UR-1116 (1989).
\bibitem{kota2}J.B. French, V.K.B. Kota, A. Pandey and S. Tomsovic, 
Ann. Phys. NY {\bf 181}, 198 (1988); Ann. Phys. NY {\bf 181}, 235 (1988).
\bibitem{kar1}V.K.B. Kota and K. Kar, Pramana-J. Phys. {\bf 32}, 647 (1989).
\bibitem{majumdar1} D. Majumdar, {\it Ph.D. Thesis}, The M.S. University of
Baroda, India, 1994 (unpublished).
\bibitem{majumdar2} V.K.B. Kota and D. Majumdar, Nucl Phys. A {\bf 604}, 129 
(1995).
\bibitem{kota3}J.B. French, V.K.B. Kota, A. Pandey and S. Tomsovic, Phys.
Rev. Lett. {\bf 58}, 2400 (1987). 
\bibitem{majumdar3} V.K.B. Kota and D. Majumdar, Z. Phys. A {\bf 351}, 
365 (1995).
\bibitem{gdr}D. Majumdar, K. Kar and A. Ansari, J. Phys. G {\bf 24}, 
2103 (1998).
\bibitem{khm}V.K.B. Kota, D. Majumdar, R. Haq and R.J. Leclair, Can. J. Phys.
{\bf 77}, 893 (1999)
\bibitem{ho83} Proceedings of the international conference on 
{\em Electromagnetic Properties of Atomic Nuclei}, Ed. H. Honie and 
H. Ohuma (Tokyo Institute of Technology Press, Japan) 1983.
\bibitem{br77} P.J. Brussaard and P.W.M. Glaudemans, {\em Shell Model 
Application in Nuclear Spectroscopy} (North Holland, New York) 1977.
\bibitem{murphy} M.J. Murphy, Ap. J. Supp. {\bf 42}, 385 (1980).
\bibitem{sc83} G.K. Schenter and P. Vogel, Nucl. Sc.. Eng. {\bf 83}, 393 (1983).
\bibitem{hillman}M. Hillman and J.R. Grover, Phys. Rev. {\bf 185}, 1303 (1969).
\bibitem{baranger}M. Baranger and K. Kumar, Nucl. Phys. A {\bf 110}, 490 (1968);
D.R. Bes and R.M. Sorensen, Adv. Nucl. Phys. {\bf 2}, 129 (1969); S. Aberg,
Phys. Lett. B {\bf 157}, 9 (1985).
\bibitem{schatz}H. Schatz  et al., Phys. Rep. {\bf 294}, 167 (1998).
\bibitem{dilg}W. Dilg, W. Schantal, H. Vonach and M. Uhl, Nucl. Phys. A
{\bf 217}, 269 (1973).
\bibitem{kas}K. Kar, S. Chakravarti, A Ray and S. Sarkar, J. Phys. G {\bf 24},
1641 (1998); S. Chakravarti, K. Kar, A. Ray and S. Sarkar, 
e-print no. astro-ph/9910058. 
\end{thebibliography}
\end{document}